\documentclass[apj, singlecolumn]{aastex63}
\usepackage{epstopdf}
\usepackage{graphicx}
\usepackage{wrapfig}
\usepackage[utf8]{inputenc}
\usepackage{amsmath}
\usepackage{natbib}
\usepackage{multirow}
\usepackage{longtable}
\bibpunct{(}{)}{;}{a}{}{,}
\usepackage{comment}
\usepackage{longtable}
\usepackage{color}
\definecolor{darkgreen}{RGB}{0,142,128}
\definecolor{darkblue}{RGB}{0,100,170}
\definecolor{darkpurple}{RGB}{150,0,150}

\usepackage{hyperref}
\hypersetup{colorlinks,citecolor=blue,linkcolor=blue}

\begin{document}
\title{Exoplanet Radio Transits as a Probe for Exoplanetary Magnetic Fields - Time-dependent MHD Simulations}
%Exoplanets at Radio, UV and X-ray Wavelength
\author[0000-0002-7069-1711]{Soumitra Hazra}
\email{soumitra\_hazra@uml.edu, soumitra.hazra@gmail.com}
\affiliation{Lowell Center for Space Science and Technology, University of Massachusetts Lowell, 600 Suffolk Street, Lowell, MA 01854, USA}
\author[0000-0003-3721-0215]{Ofer Cohen}
\affiliation{Lowell Center for Space Science and Technology, University of Massachusetts Lowell, 600 Suffolk Street, Lowell, MA 01854, USA}
\email{ofer\_cohen@uml.edu}
\author[0000-0002-6118-0469]{Igor V. Sokolov}
\affiliation{Department of Climate and Space Sciences and Engineering, University of Michigan, 2455 Hayward St., Ann Arbor, MI 48109, USA}
\email{igorsok@umich.edu}

\begin{abstract}
We perform a series of time dependent Magnetohydrodynamic simulations of the HD 189733 star-planet system in order to predict radio transit modulations due to the interaction between the stellar wind and planetary magnetic field. The simulation combines a model for the stellar corona and wind with an exoplanet that is orbiting the star in a fully dynamic, time-dependent manner. Our simulations generate synthetic radio images that enable to obtain synthetic radio lightcurves in different frequencies. We find a clear evidence for the planetary motion in the radio lightcurves. Moreover, we find specific repeated features in the lightcurves that are attributed to the passage of the planetary magnetosphere in front of the star during transit. More importantly, we find a clear dependence in magnitude and phase of these lightcurve features on the strength of the planetary magnetic field. Our work demonstrates that if radio transits could be observed, they could indeed provide information about the magnetic field strength of the transiting exoplanet. Future work to parameterize these lightcurve features and their dependence on the planetary field strength would provide tools to search for these features in radio observations datasets. As we only consider the thermal radio emission from the host star for our study, very sensitive radio interferometers are necessary to detect these kind of planetary transit in radio.

\end{abstract}


\section{Introduction}
Since the discovery of first planet outside the solar system, thousands of exoplanets have been confirmed \citep{mayo95a, schi95a}. Following the dedicated {\it Kepler} \citep{hans10a} and {\it TESS} \citep{RickerTESS} missions, we now have significant statistical information regarding the masses, sizes, orbital separations of these transiting exoplanets. Specifically, many of these exoplanets are found in a short-period orbit, with a semi-major axis of less than 0.1 AU (sometimes even less than 10 stellar radii) \citep{schi95a}. Most of these close-in exoplanets are hot gas giants known as hot Jupiters, which are expected to produce a strong star-planet interaction due to their close-in orbit \citep[sometimes located within the Alfv\'en radius, see, e.g.,][]{shko03,Ip2004,Lanza2008,cohe11,cohe18}. Indeed, Many observations regarding close-in star-planet system have been reported using modern space based telescope and ground based instruments \citep[see summary in][]{stru18a}. 

Exoplanet observations are now strongly supplemented by the consistent growing observational efforts of detecting spectral emission from the atmosphere of these exoplanets. These include observations of Lyman-$\alpha$ signature of the atmospheric evaporation \citep{vida03, trip15, bour16, spak18, vido21} and chromospheric signature of the star-planet interaction \citep{shko05, shko08, fare10, shko18}. However, consistent observational techniques to detect exoplanetary magnetic fields are still missing. Magnetic fields may play a crucial role in the planetary evolution, they may (or not) protect exoplanets atmospheres, and may provide an insight about exoplanets internal structure \citep[see, e.g.,][]{ExoplanetHandbook}. Thus, exoplanets magnetic field observations are crucial for exoplanets characterization. In close-in exoplanets, star-planet interaction observations in the Radio, EUV and X-Ray may provide an insight about the planetary magnetic field \citep{zark07a, bent17a, grie18, zark18, stru22}.

In our solar system, radio signals from Jupiter have been observed \citep{burk55}. Thus, we could pose the question is it possible to detect exoplanets by looking at radio signals? Since the first discovery of radio emission from Jupiter, similar type of radio emissions are also detected for other planets in the solar system \citep{galo89}. It is now well known that planetary magnetosphere extract energy from the solar/stellar wind of the host star and some part of this energy can be radiated via electron cyclotron maser instability, likely at radio frequencies \citep{Gurnett1974,Desch1984,grie07, lazi07, lazi18, lync18, zark18}. Depending on the magnetic nature of the host star and the planet, four types of interaction between the stellar wind and planet is possible. It has been shown that three of these four possible cases intense radio emission is possible \citep{zark07a}. Intense radio emission is not possible only when both star and the planet are non-magnetic. 

Similar like solar system planets, intense cyclotron maser emission at radio wavelength have been predicted for hot Jupiter exoplanets. Detection of these radio emission using ground based instrument is limited by the ionospheric cut-off frequency, approximately around 10 MHz \citep{davi69, yeh82}. Solar system planets except Jupiter emit very low radio frequency below this cutoff, making detection very difficult from ground based present instrument. This is due to very weak planetary magnetic field strength, as frequency of the radio emission is directly proportional to the magnetic field strength close to the planetary surface \citep{grie18}. Low frequency ($\leq 200 MHz$) radio emission is also one of the known tools to probe the outer stellar corona and space weather conditions around that star \citep{schw06, vedan20a}. In summary, signature of the star-planet interaction can be detected either by observing the planetary auroral emission or by observing the modulation in the planetary radio transit. This realization makes space-based or ground-based very low frequency radio observatory as a promising candidate for the detection of exoplanet in radio band \citep{burkh17, grie18, zark18, pope19}. 

Until recently, neither the exoplanet nor their host star have been detected in the low frequency radio band despite several try \citep{lync18}. Thanks to the higher sensitivity of the new generation radio interferometer LOFAR \citep[the LOw-Frequency ARray;][] {vanh13}, coherent low-frequency radio emission from M dwarf stars have been detected recently \citep{vedan20, calli21a, calli21}. Radio emissions are also detected for some other stellar systems using LOFAR \citep{turn21}. It has been suggested that this radio emissions are similar to that of planetary auroral emission, indicating the signature of star-planet interaction via electron cyclotron maser instability \citep{vedan20a, calli21}. In the case of M dwarf star GJ 1151, although coherent radio emissions were detected, no conclusive evidence of the massive planet around that star was found \citep{pope20a, maha21,perg21, pope21}. Several theoretical studies have been published regarding exoplanet detection in the radio band \citep{zark97, zark07a, selho13, vido15, cohe18, kavan19, kavan20, selho20}. \cite{selho13, selho20} studied the possibility of detecting exoplanets at high radio frequencies (17 GHz and more than that) using planetary transit and suggest that it is possible to observe this kind of planetary transit with the Atacama Large Millimeter/Submillimeter Array (ALMA) radio interferometer. Observational facilities like LOFAR, the upgraded version of the existing Giant Meter Radiowave Telescope \citep[GMRT;][] {gupt17} and Murchison Widefield Array \citep[MWA;][] {ting13} actually opens up the unique opportunity for the more sensitive search in the low frequency range \citep[See][]{pope19, shioh20}. The upcoming planned Square Kilometer Array \citep[SKA;][] {dewd09} is also expected to conduct a very sensitive search in low frequency range which will be helpful to find out the signature of star-planet interaction. \cite{pope19} calculated the radiometric sensitivity of the upcoming SKA and suggest that it is possible to detect close-orbit exoplanet transit around the host star using SKA. Motivated by the present observational success as well as theoretical studies, here we aim to study the possibility of exoplanet detection and characterization in the radio band from the MHD modeling point of view.

In this study, we follow the approach suggested by \cite{cohe18} for the detection and characterization of exoplanets' magnetic fields via the planet-induced modulation of the background coronal radio emission, instead of detecting the planet as a radio source. \cite{cohe18} mimicked the orbital phase variation of the exoplanet around the host star by viewing the static, three dimensional solution from different angles. However, this method missed the variation of plasma properties along the planetary orbit when the planet actually moves. Here we perform the time-dependent simulation of the star-planet interaction to study similar, but dynamical system. Basically, we aim to use the planetary transit for the characterization of radio emission from the host star. That will also help us to detect and characterize exoplanets. Please note that in this study we only focuses on the thermal radio emission from the host star, not the coherent radio emission. Coherent radio emission generally comes from small regions of the star that can be very strongly lensed and time variable.

We describe the details of our model in Section~\ref{Model}. We present the details of our results in Section~\ref{Results} and a detailed discussion of our results in Section~\ref{Discussion}. Our method of exoplanet magnetic field measurement from the radio transit are described in Section~\ref{exo-mag}. Section~\ref{uv-xray} describes our results regarding exoplanet detection in UV and X-ray. Finally, we present a summary of this study and our conclusions in the last section.

\section{Time dependent model of the Star-Planet interaction}
\label{Model}

We developed the time-dependent model of the star-planet interaction using the BATS-R-US global MHD model \citep{powe99, toth12} and its version for the stellar corona and wind, the Alfv\'en Wave Solar Atmosphere Model \citep[AWSOM][]{vand14}\footnote{The BATS-R-US and AWSOM codes are part of the open-source Space Weather Modeling Framework (SWMF), which is available at \url{https://github.com/MSTEM-QUDA}. Input parameter setting file as well as any results data file are available upon request. The code version and the input parameter file provide the ability to fully recover the results presented here.}. AWSOM has been used extensively to study different properties of the solar corona and the solar wind. This model solves the set of non-ideal MHD equations (mass continuity, momentum, magnetic induction and energy equations) in the conservative form, while taking into account thermodynamic processes. Our time-dependent Star-Planet interaction setup consists of two parts. First we simulate the ambient solar/stellar wind, and second, we superimpose the planet in to this background solution.

\subsection{Modeling Stellar Corona}

In the AWSoM set up, the propagation, reflection, and dissipation of the Alfv\'en wave energy are modeled by solving two additional equations - one for waves propagating parallel to the magnetic field and another is for waves antiparallel to the magnetic field. We refer the reader to \cite{vand14} for the complete detailed description of the model. In the AWSoM formalism, the Alfv\'en wave pressure gradient accelerates the solar wind plasma \citep{alaz71}. Non-linear interaction between the outward and counter-propagating Alfv\'en waves generates turbulent cascade which is the source for coronal heating \citep{tu93, tu95,chan09}. Detailed thermodynamic effects, such as radiative cooling and thermal conduction are also included in the AWSoM setup. We use Threaded Field Line Model (TFLM) to model the transition region and lower corona as prescribed by \cite{soko21}. This helps to save the computational resources which otherwise needed to resolve the fine structure of the transition region using a highly refined grid. We refer the readers to \cite{sach19,sach21} for the validation study of the AWSoM model with observations.

One can develop the model for the solar wind by initializing the AWSoM model with the solutions from a Potential Field Source Surface (PFSS) magnetic field extrapolations \citep{scha69} obtained from the synoptic maps of the solar photospheric radial magnetic field. To develop the model for stellar wind, one have to just use the photospheric radial maps of that specific star instead of the Sun. Thanks to Zeeman-Doppler imaging techniques, these kind of observations for stars other than the Sun are available \citep[e.g.,][]{dona97, dona99}. For this study, we use the HD189733 stellar system. The stellar parameters of HD189733 (stellar mass M$_*$, stellar radius R$_*$ and stellar rotation period P$_*$) are listed in the Table \ref{tab:my_label}, and the model is driven by magnetogram data obtained from \cite{fare10}. Using AWSoM, we obtain the self-consistent, steady-state solution of the stellar corona and stellar wind. 
\begin{table}[]
   \caption{Stellar Parameters of HD 189733}
    \centering
    \begin{tabular}{c c}
    \hline \hline
    Stellar Parameter   ~~~~~ ~~~~&   Value\\
    \hline
      R$_*$   &  0.76 R$_\odot$\\
      M$_*$   &  0.82 M$_\odot$\\
      P$_*$   &  11.95 days    \\
      \hline
    \end{tabular}
    
    \label{tab:my_label}
\end{table}

\subsection{Modeling the Planets}
In our star-planet simulation, the planet is modeled through an additional boundary conditions for a second body that is imposed in the simulation domain. In our setup, the second body is the planet and the first body is the star. Next, we aim to include the orbital motion of the second body in our model. For that purpose, we updated the coordinates of the second body along a circular orbit with a radius of the planet's semi-major axis. In principle, any kind of orbit is possible, but for the sake of simplicity here we assume a circular orbit in the equatorial plane. In future work, we plan to generalize the planetary orbit to include additional orbital parameters, such as inclination and eccentricity.

To develop the time dependent model of the star-planet interaction, we first determine the cells which are inside the second body. We define the cells inside the second body as "body cells" and cells outside the body and the first body (star) as "true cells". Next, we update the coordinates of the second body per the orbital motion of the planet. When we update the second body coordinate, some previously body cells now become true cells as they are now outside the second body. On the other hand, some previously true cells now become body cells. New true cells, which were inside the second body before, need to be filled. We fill these cells by the averaging of nearest surrounding true cell values. As BATS-R-US uses block adaptive techniques, one have to also update ghost cells if the new true cells are near the block boundary. Cells that were true cells and are now body cells, are filled with the boundary conditions for the second body. In essence, this procedure dynamically move the boundary conditions of the second body along the planetary orbit.

Please note that frequency of the second body coordinate update is constrained by the numerical stability condition and the timestep. It is also necessary to resolve the second body well; for that purpose we need very fine grid resolutions (at least 10 grid cells across the planetary body). We follow the prescription of \cite{cohe11, cohe18} for that purpose. We set the planet size as $0.2 R_\odot$, which is almost twice size of HD 189733 b. \cite{cohe18} have shown that model results are independent of the planet size up to $0.15 R_\odot$. However, very fine resolution is needed to resolve the smaller size planet, making the computation very expensive. We follow grid refinement strategy along the planetary trajectory to resolve the planet well at any point of its trajectory. One can see \cite{cohe11, cohe18} for further details.

As the magnetic field strength of the planet is not known, we use two specific cases for our simulations. Because of tidal locking and larger rotation period, one can expect lower magnetic field for hot Jupiter planets compared to that of Jupiter \citep{sanc04}. However, stronger planetary magnetic field is also necessary to protect the planetary atmosphere from the erosion by the stellar wind. Keeping these in mind, we formulate our cases. In one case, we set the higher planetary magnetic field (3 G) and in another case lower Earth-like planetary magnetic field (0.3 G). We also consider a non-magnetized case.  For our simulation, we use planetary boundary number density as $10^7$ cm$^{-3}$  and boundary temperature value as $10^4$ K. These values are sufficient to produce significant modulation in the background coronal density, although it produces lower thermal outflow from the planet compared to hot Jupiter \citep{cohe18}. Increase in these values will only intensify the modulations. For this study, we also consider two different orbital separations. In one case, we place the planet at a distance of $10 ~R_*$ from the star (a short-orbit case), and in another scenario, we place the planet at a distance of $20 ~R_*$ from the central star (a longer-orbit case). $R_\star$ is the stellar radius.

\subsection{Synthetic Images of the Stellar Corona in Radio, UV and X-ray Wavelength}

\subsubsection{Synthetic Radio Images}

We use the utility presented in \cite{mosc18} for generating synthetic radio images of the stellar corona from our MHD wind solutions. This algorithm captures the process of radio emission from the stellar corona due to Bremstrahulong emission and the way it propagates through the circumstellar medium of the non-uniform density \citep{benk10, benk12, mosc18}. During the propagation, radio waves suffer refraction. Although all the electromagnetic waves face the effect of refraction, radio waves experience more refraction because of their strongly varying refractive index between the medium of different densities \citep{kund65, ober11,moha17}. 

Our radio emission calculation tool uses the ray tracing algorithm developed by \cite{benk10}, to calculate the actual curved path for the different frequency radio rays. Radio wave refraction actually controls the curved path trajectory for a given frequency $\nu$ (and angular frequency, $\omega$) from one grid cell to another inside the computational domain. Refractive index is related with the dielectric permittivity $\epsilon$ via the dispersion relation:\\
\begin{equation}
    n^2= \epsilon= 1- \frac{\omega_p^2}{\omega^2}
\end{equation}

where $\omega_p=4 \pi e^2 n_e/m_e$ is the plasma frequency, with the electron number density, $n_e$, the electron mass, $m_e$, and the electron charge, $e$. Dispersion relation indicates that if the plasma frequency is greater than the ray frequency, refractive index become imaginary. Dispersion relation also suggests that in a region where plasma frequency is less than the ray frequency, low radio frequencies suffer increased refraction compared to higher frequencies (e.g., optical, EUV, and X-ray). One can assume the quasi-neutrality of the plasma and write the hydrogen plasma density as $\rho=m_p n_e$.  We can then rewrite the dispersion relation as:
\begin{equation}
    n^2=\epsilon=1- \frac{\rho}{\rho_{cr}}
\end{equation}
where $\rho_{cr}$ is the critical plasma density where refractive index becomes zero and radio waves can not transmit. 

Finally, this tool calculates the radio intensity by performing an integration over the ray trajectories and provides a radio image for the particular radio frequency. We calculate the intensity of each pixel, $I_\nu$, by performing the integration over the emissivity along each ray path for a particular frequency $\nu$:

\begin{equation}
    I_\nu=\int B_\nu(T) k_\nu ds
\end{equation}

where Bremsstrahlung emission are represented by the term:
\begin{equation}
    B_\nu (T)=\frac{2 k_B T_e \nu^2}{c^2}
\end{equation}
and the absorption coefficient is: 
\begin{equation}
    k_\nu= \frac{n_e^2 e^6}{\nu^2 (k_B T_e)^{3/2}m_e^{3/2}c} <g_{eff}>
\end{equation}

Here, $k_B$ is the Boltzman constant, e is the electron charge, $n_e$ is the electron number density, $T_e$ is the electron temperature, $m_e$ is the mass of the electron, c is the speed of light and $<g_{eff}>$ is the Gaunt factor which is assumed to be equal to 10 in our study \citep{karz61}. 

\subsubsection{Synthetic UV and X-Ray Images}
We generate synthetic UV and X-ray images by performing a line of sight integration:
\begin{equation}
    I_{Pix}=\int n_e^2 \Lambda (T) ds,
\end{equation}
where, $I_{Pix}$ is the flux in each pixel, $n_e$ is the total electron density, $\Lambda (T)$ is the temperature response function obtained from the CHIANTI database \citep{Landi2012} and $ds$ represents the differential path along the line of sight. Finally, we generate synthetic light curve using the generated synthetic UV and X-ray images for different planetary phases.

\begin{figure*}[h!]
\centering
\includegraphics[width=6.75in]{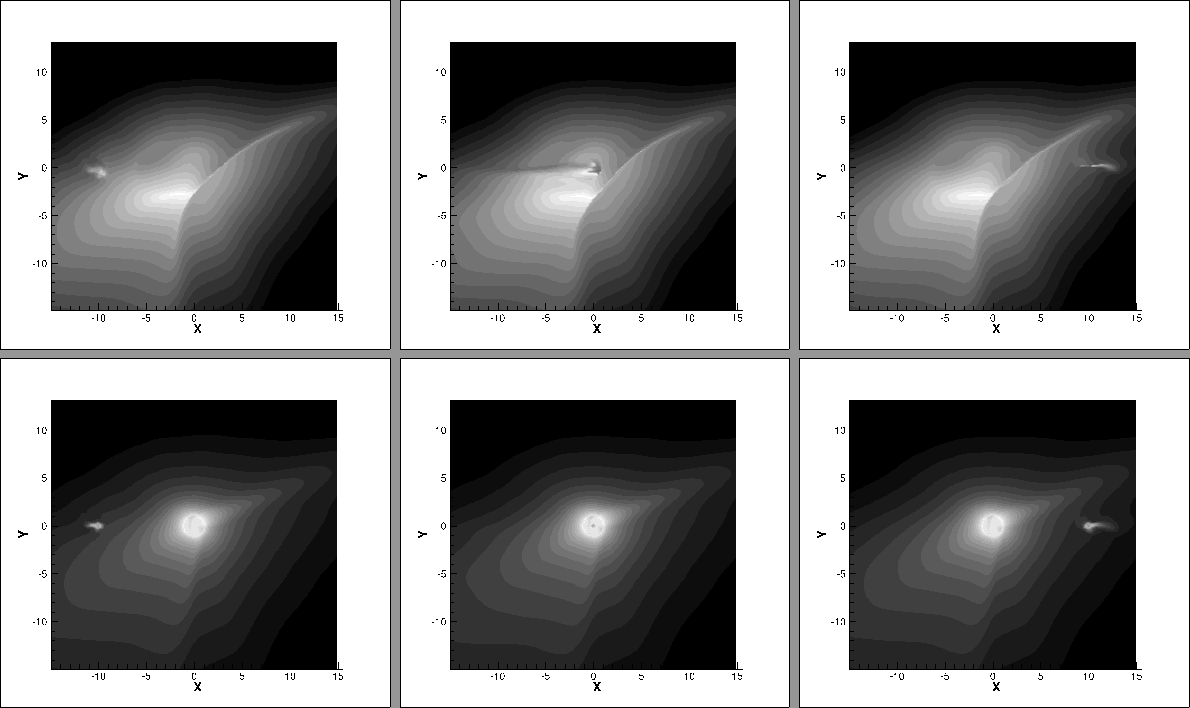}
\caption{A sample of synthetic radio images produced by the model. Results are shown for the 10 MHz (top) and 1 GHz (bottom) for the corresponding viewing phase angle of 0.25 (left), 0.5 (middle), and 0.75 (right). These images are for the case where the planet is located at 10R$_*$ and a planetary field of 3 G, and the planet's signature is clearly seen in all images. The planet is visible in the radio images for all other cases as well.}
\label{fig:RadioImage}
\end{figure*}

\begin{figure*}[!ht]
\centering
%\begin{tabular}{cc}
\includegraphics*[width=1.0\linewidth]{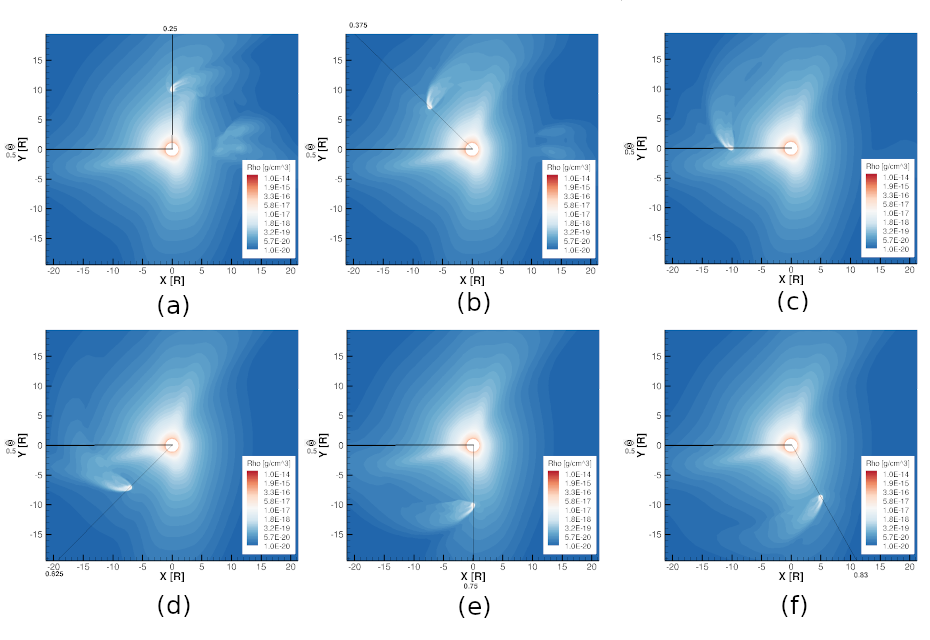} \\
\caption{Evolution of the star-planet interactions at different planetary phase. Thin black line corresponds to different planetary phases like 0.25, 0.375, 0.458, 0.5, 0.625 and 0.75. Thick black line corresponds to the mid-transit line (phase 0.5) along which the telescope is placed. Eye symbol in the figure represents the observing telescope. The planet is placed at a distance of 10R$_*$ from the star.}
\label{fig:evol-short}
\includegraphics*[width=1.0\linewidth]{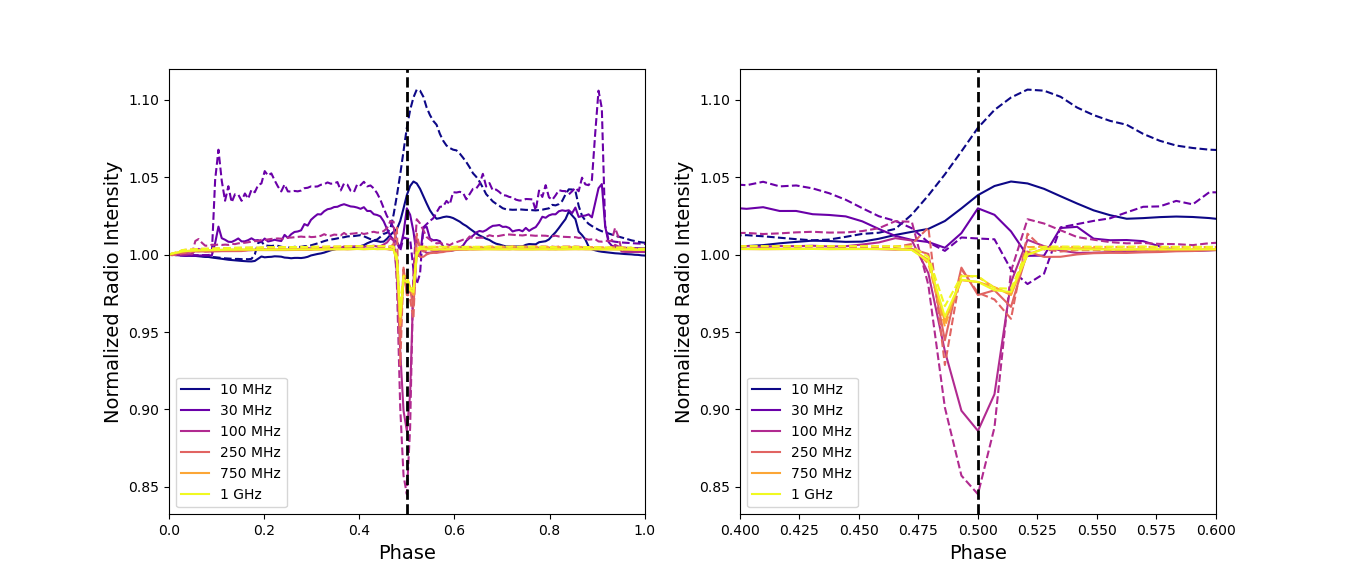}
%\end{tabular}                
\caption{\footnotesize{Left panel shows the synthetic light curves of the different radio frequencies (shown in different colors and line-style) for two different cases. In one cases, we take the planetary magnetic field as 3 G (dashed curve H in the figure) and in another case, we take the value as 0.33 G (solid curve in the figure). Thick black dashed line indicates the mid point of the transit. Right panel shows the same but zoomed in, focuses only the transit phase between 0.4 and 0.6. Planet is placed at a distance of 10R$_*$ from the star.}}
\label{fig:exo_short_transit1}
\end{figure*}

\begin{figure*}[h!]
\centering
\includegraphics[width=6.75in]{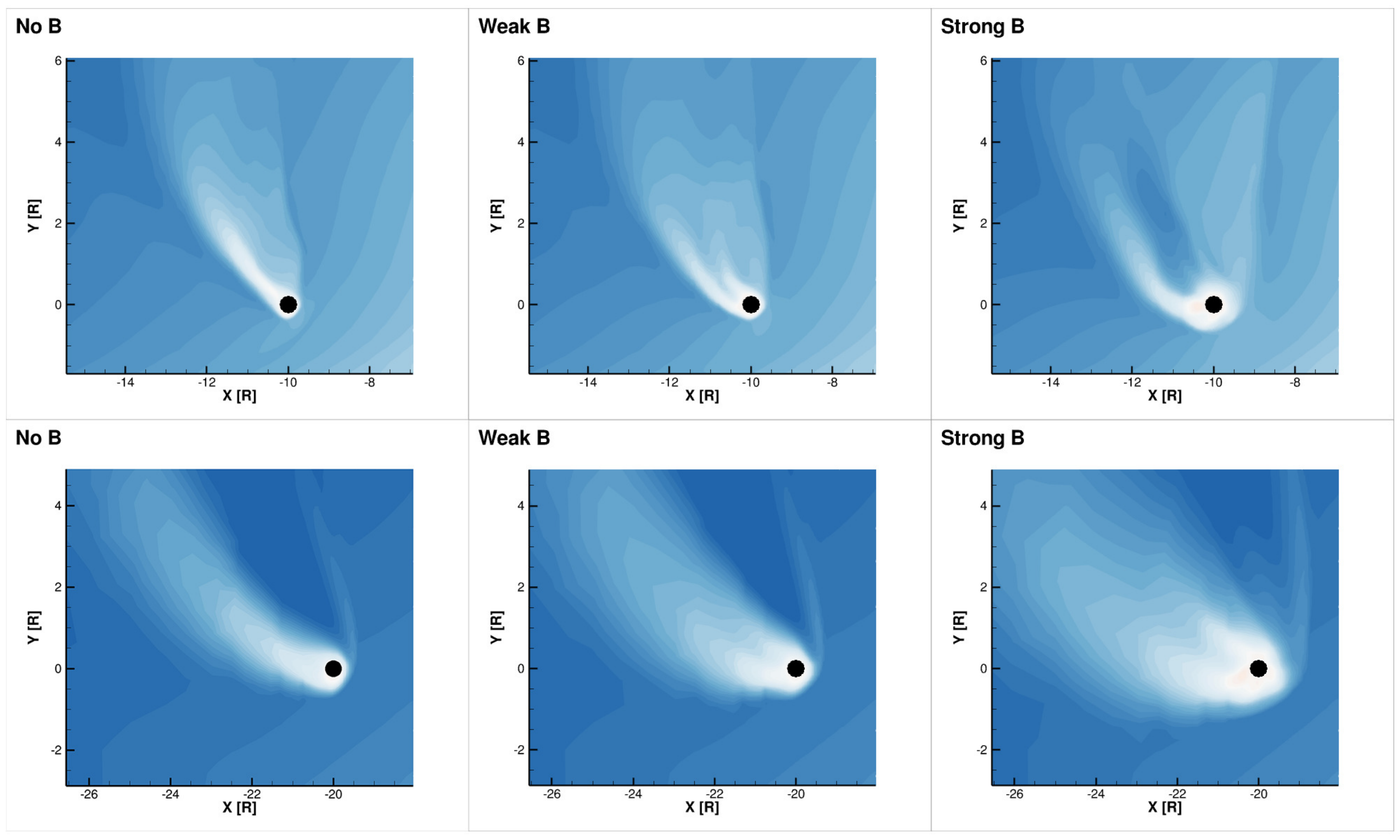}
\caption{Top: a zoom on the planetary magnetosphere for the close-orbit case for a non (left), weakly (middle), and strongly (right) magnetized planet. Bottom: a zoom on the planetary magnetosphere for the far-orbit case for a non (left), weakly (middle), and strongly (right) magnetized planet. Display is similar to Figure~\ref{fig:evol-short}. This figure indicates that the size of the planetary magnetosphere is larger for the higher planetary magnetic field.}
\label{fig:Magnetosphere}
\end{figure*}

\begin{figure*}[!ht]
\centering
%\begin{tabular}{cc}
\includegraphics*[width=1.0\linewidth]{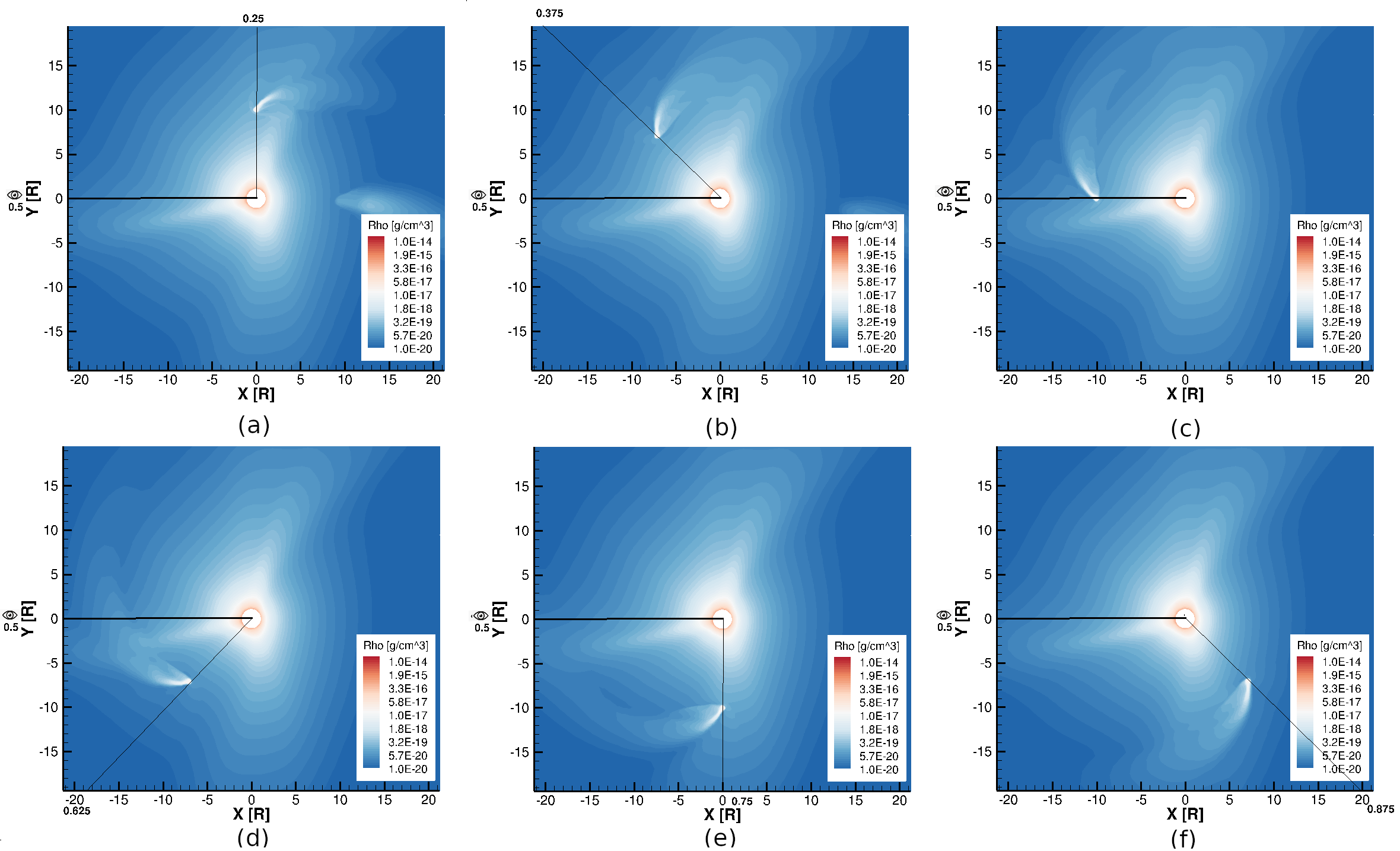}\\
\caption{Same as Figure \ref{fig:evol-short} but for the case where the planet is nonmagnetized. Thin black line corresponds to different planetary phases like 0.25, 0.375, 0.5, 0.625, 0.75 and 0.875. Thick black line corresponds to the mid-transit line (phase 0.5) along which the telescope (eye symbol in the figure) is placed.}
\label{fig:evol-unipolar}
\includegraphics*[width=1.0\linewidth]{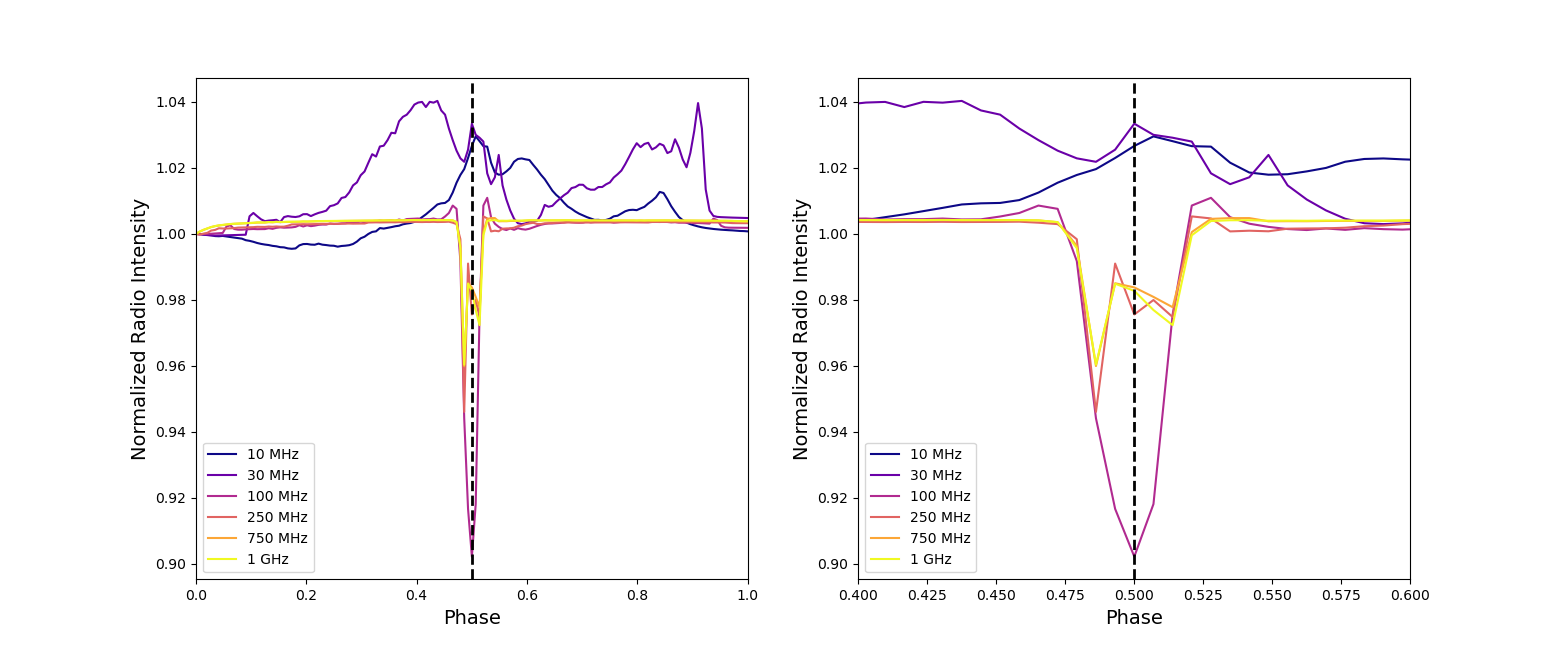}
%\end{tabular} 
\caption{Same as Figure \ref{fig:exo_short_transit1} but for the case when the planet is nonmagnetized. Thick black dashed line corresponds to the mid-transit line (phase 0.5).}
\label{fig:exo_uni_transit}\
\end{figure*}

\begin{figure*}[!ht]
\centering
%\begin{tabular}{cc}
\includegraphics*[width=1.0\linewidth]{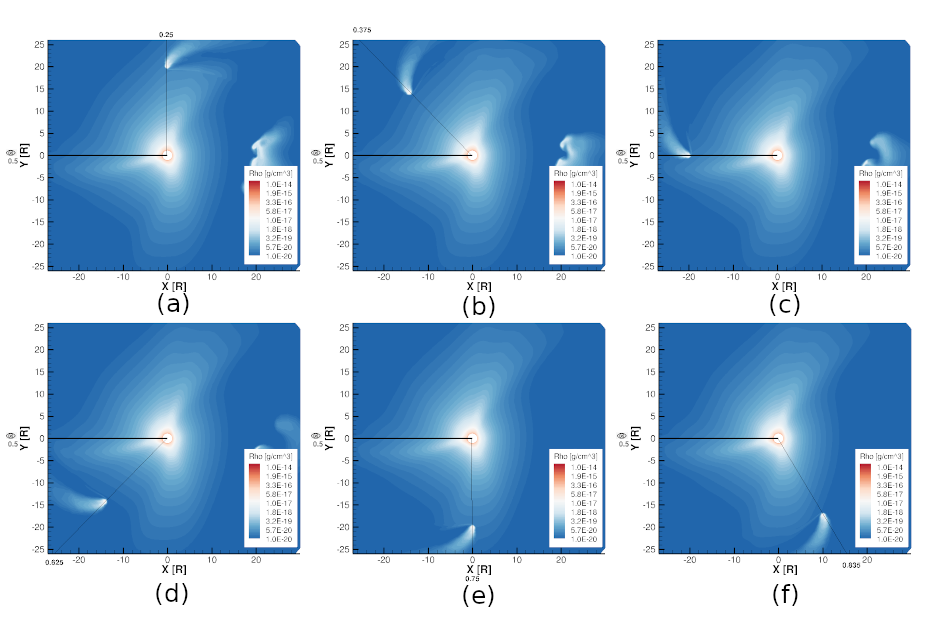}\\
\caption{Same as Figure \ref{fig:evol-short} but for the case where the planet is located at 20 $R_*$ . Thin black line corresponds to different planetary phases like 0.25, 0.375, 0.5, 0.625, 0.79 and 0.83. Thick black line corresponds to the mid-transit line (phase 0.5) along which the telescope (eye sign) is placed.}
\label{fig:evol-long}
\includegraphics*[width=1.0\linewidth]{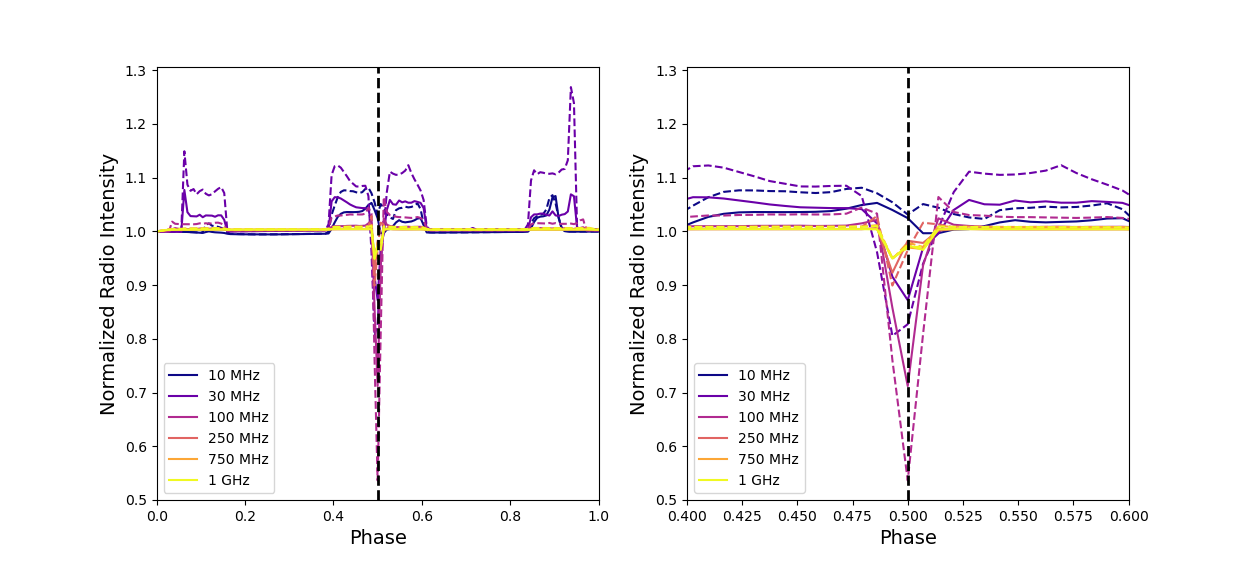}
%\end{tabular} 
\caption{Same as Figure \ref{fig:exo_short_transit1} but for the case when the planet is placed at 20 $R_*$ from the central star. Thick black dashed line corresponds to the mid-transit line (phase 0.5). Dashed line corresponds to the higher magnetic field (3 G), while solid line corresponds to the lower magnetic field (0.33 G).}
\label{fig:exo_long_transit1}
\end{figure*}

\section{Results}
\label{Results}

In this section, we aim to characterize radio transit signals. In each case, we first perform a simulation keeping the planet fixed at a certain distance and obtain the steady state solution for the stellar corona and stellar wind, including the planet. This way, the wind and coronal solution evolves undisturbed by the planetary motion. Next, we use that steady state solution as an initial condition and perform the time dependent simulation where planet moves along the orbit. This way, we capture the disturbance of the coronal solution by the moving exoplanet. In this study, a hot Jupiter planet is moving around the HD 189733 star. The choice of an actual star over an idealized stellar dipole field is to capture a more realistic (non-uniform) background stellar radio emission. Our simulation captures the variation of plasma density along the planetary orbit as the planet moves. We generate the synthetic radio images of the stellar corona at different phases of the planetary orbit and calculate the modulation of this coronal radio emission by the Hot Jupiter exoplanet. Here, we consider the semi-major axis and the planetary magnetic field strength as our two free parameters. Other parameters, such as the planetary field orientation and planetary rotation, are left for a future study. 

Our synthetic observations assume a single observing point connecting the observer and the star, while the exoplanet is orbiting around the star, modulating the medium along the line-of-sight (LOS) as a function of the orbital phase. The moving exoplanet can modulate the radio emission in two ways. In can block or deflect some of the emission from the observer, leading to an overall flux reduction, and it can focus the emission towards the observer, leading to an overall increase in the observed flux. These two effects and their impact on the observed radio lightcurves are discussed for each case.

We would like to emphasis that we intentionally do not scale the flux to certain distance from the Earth, but we only show the modulations observed by a "local" observer. The assumption we make in calculating synthetic modulations of the ambient radio flux is that the stellar flux is actually observable from the Earth. By definition, this work is only relevant to stars with an observable flux from the Earth, so we do not provide the actual flux magnitude in units of Jy. \cite{cohe18} have listed a number of stars with realistically observable radio flux, arguing that the methodology presented here is useful.

\subsection{Short-orbit Case}
\label{shortO}

We first consider the case when the planet is placed at a distance of 10 stellar radii ($a=0.037$AU). In this case, the planetary magnetosphere is sweeping through a relatively high background coronal density and magnetic field. Thus, strong radio modulations are expected. Indeed, the radio intensity in different frequency bands is strongly modulated by the star-planet interaction as the planet is very close to the host star. When radio waves at these frequencies propagate through the ambient medium between the star and the planet, they suffer a strong refraction due to the strong density variations between the ambient corona and the planetary magnetosphere. The magnetopause region, where plasma is highly compressed also plays a role in the modulations of the radio path. Figure \ref{fig:RadioImage} shows synthetic radio images for frequencies of 10 MHz and 1 GHz respectively. Modulation of the ambient coronal plasma by the planetary movement are clearly reflected in these generated radio images.

The interaction described above is shown in Figure \ref{fig:evol-short}. The figure shows a top-down view on the equatorial plane, colored with density contours. It shows the density modulation of the ambient medium by the moving planet for different planetary phases, where the planetary magnetosphere is clearly visible. The telescope position and the LOS are marked by the thick black line (along the Y=0 line in the negative x direction), so the centre transit point (phase 0.5) is given by this point of observation. When the planet is at phase 0.25 (Fig. \ref{fig:evol-short} a), the coronal density along the LOS is affected slightly by the planetary movement. We also note that the planetary magnetosphere is slightly tilted from the normal planetary phase line (thin black line) due to the planetary orbital motion (leading to a comet-like magnetotail). When the planet moves close to the mid transit line, density modulations starts to impact the collection of radio waves by the telescope. This is basically due to radio frequency refraction which is controlled by the density modulation of the ambient medium. Even if the planet moves away from the mid-transit line, the tail of the planetary magnetosphere still has some impact on the density of the ambient medium along the LOS (see Fig. \ref{fig:evol-short} d-f). We also note again that previous studies mimicked the orbital phase variation of the exoplanet around the host star by viewing the static solution from different angles \citep[a moving observer, static soution, e.g.,][]{cohe18,stru18a}. Please note that there is some initial density perturbation at the start of the simulation (phase zero); but that will not impact the result as that phase is on the complete opposite side of the line of sight.

Figure \ref{fig:exo_short_transit1} shows the synthetic light curves of the different frequency radio intensity as a function of the orbital phase for two different planetary magnetic field strengths. Radio intensity values are presented in relative flux values normalized to the flux at phase zero (when the planet is eclipsed by the star). %Radio intensity values are presented in relative flux values normalized to the maximum flux value at each frequency. 
Our virtual observing telescope is located in the middle of each plot (at a phase of $0.5$) designated as a mid-transit point. Our "telescope" is located at $40~R_*$ along the mid-transit point. The solid and dashed synthetic light curves in Figure \ref{fig:exo_short_transit1} correspond to two different planetary magnetic field strengths of 0.33 Gauss and 3 Gauss. Both these cases show similar trends, but they differ in magnitude and phase (see Figure \ref{fig:exo_short_transit1}). 

Figure~\ref{fig:Magnetosphere} shows a zoom in view near the planetary magnetosphere for the three cases (strongly, weakly, and non magnetized planet). It can be seen that the day-side magnetosphere is very small in the non and weakly magnetized case (about 0.5 planetary radii), and its is larger for the strongly magnetized case (about 2.5 planetary radii). It can also be seen that the strongly magnetized magnetotail is wider and it has more structure, possibly due to a magnetic interaction with the ambient coronal plasma and magnetic field.

Figure \ref{fig:exo_short_transit1} shows that high frequency radio emission (equal to or more than 100~MHz) is blocked by the hot Jupiter planet near the mid-transit phase, causing a drop of around 5-10~\% in intensity. The higher the frequency is, the lower the drop we see. This high frequency blocking trend follows the logic that the high frequency emissions come from the hot, dense regions at the low corona where the magnetic field is stronger \citep[e.g.,][]{cohe18,mosc18}, which are simply shaded by the planet. We notice a slight increase in the intensity when planet starts to move out or in of transit. It seems like as the planet is moving out of transit, and the magnetotail is located at the transit point, there is an overall focusing of the radio wave path at high frequencies, leading to a flux increase. When planet starts to enter into the transit phase, then also some refracted waves reached into the telescope, resulting an increase in flux just before the transit. We found that drop in the radio intensity is higher (around 15-20~\%) for 100~MHz frequency. The higher intensity drop at 100~MHz-1~GHz, mid-transit for the higher field strength indicates a larger flux blocking. This makes sense if one assumes a slightly larger magnetosphere for a higher field strength (Figure~\ref{fig:Magnetosphere}), which blocks a larger area of the background emission.

The behaviour of radio transit is quite different in the low-frequency bands (below 100 MHz). The lower frequency stellar radio emission is associated with the cooler, less dense regions of the higher corona, which are also the regions that the planet is passing through. Transit behaviour of the 30~MHz radio frequency (Figure~\ref{fig:exo_short_transit1}) indicates an increase in the lower frequency radio intensity starting around phase 0.2 and then drops at the mid transit phase. Drop in the mid transit phase is expected due to the flux blocking by the planetary magnetosphere. We also note that modulation in the radio transit behaviour is more for higher planetary magnetic field scenario.

However, the behaviour of 10 MHz radio frequency transit (Figure~\ref{fig:exo_short_transit1}) is very different. It shows an increase in the radio intensity starting around phase 0.2, and increases to very high value (10~\% increase) around the mid transit phase and then starts dropping. Looking at Figure~\ref{fig:exo_short_transit1}, these phases seemed to be the locations of the dense stellar helmet streamers. Thus, the increase of the radio flux at very low frequency (10~MHz) is associated with the crossing of the denser regions of the stellar corona, possibly leading to a compression of the planetary magnetosphere, and a 8-10~\% increase in the radio waves flux that is collected by the observing telescope (see Figure \ref{fig:exo_short_transit1}). The increase at mid-transit seems to be a combination of the helmet streamer crossing (regions around the largest closed stellar magnetic field loops), and the magnetotail focusing effect also seen at low frequencies. Here, the flux increase is greater for the strong field case comparing to the weaker field case. It seems like the magnetotail size contributes more to the enhancement of the low-frequency radio flux as it crosses the observing LOS. Probably, radio frequency of 100~MHz or less than that is the best suitable frequency for observing planetary transit in radio. Finally, no significant phase shift is visible between the two field strength cases in the low frequencies.  

The total radio intensity for the higher frequencies, e.g., 1~GHz lies on the order of $10^{-14}~[W~m^{-2}~Hz^{-1}]$, while for a frequency of 10 MHz, it is on the order of $10^{-17}~[W~m^{-2}~Hz^{-1}]$. Although low frequency radio intensities modulated significantly due to star-planet interaction, the total low frequency radio emission is much lower compared to higher radio frequency emission. This makes the observation of low frequency radio emission a challenging task compared to high frequency radio observation. 

In addition to the strong/weak field cases, we also consider a non-magnetized planet case. The interaction between the magnetized stellar wind and the non-magnetized obstacle (planet) has been recently recently defined as "unipolar" interaction \citep{stru15, stru18a}. In our solar system, interaction between the solar wind and the Venus atmosphere is an ideal example of such an unipolar interaction. It was shown that the unipolar interaction leads to the creation of an induced magnetosphere having similar global structure like other self-generated planetary magnetosphere \citep[e.g.,][]{Luhmann1981,Kivelsonrussell1995,Russell2006, basa21}. However, \cite[e.g.,][]{ma13} have shown that induced magnetosphere has much less spatially extended compared to the self-generated planetary magnetoshpere.

When we perform the simulation with  a non-magnetized planet, we find an induced magnetosphere around the non-magnetized planet (see Figure \ref{fig:evol-unipolar}). This reaffirms previous findings \citep{stev03, ma13, basa21} that non-magnetized planet possess an induced magnetosphere. Figure \ref{fig:evol-unipolar} shows the density modulations of the surrounding medium due to the induced magnetosphere when the planet is non-magnetized. Figure \ref{fig:exo_uni_transit} shows synthetic light curves of the different frequency radio intensity modulations in case of the unipolar interaction scenario. We note that non-magnetized planet is also placed at a distance $10~R_\star$. The radio modulations show overall similar trends to the weakly magnetized case for higher frequencies (equal to or more than 100~MHz), but the modulations are weaker (about 10\% or less). In this scenario, we also notice higher drop in the radio intensity for 100~MHz case. For lower frequencies, we notice an increase in the radio intensity value starting around phase 0.2 and a drop around mid transit (Figure~\ref{fig:exo_uni_transit}). It seems like the particular structure of the induced magnetosphere leads to a significant focusing of the radio waves at this frequency.

Our results show that the radio waves are blocked or disrupted not only by the actual planet but also by the planetary magnetosphere. The flanks or edge of the planetary magnetosphere start disrupting the propagation of radio waves well before the beginning of the actual transit. Thus, radio emission modulations are affected more strongly by a transiting exoplanet comparing to visible transits. 

\subsection{Longer-orbit Case}
\label{LongO}

In this scenario, we place the planet at a distance of $a=20~R_\star$ ($0.080$AU). As we go outwards, the density of the stellar corona now decreases significantly. In this situation, one can assume the planetary magnetosphere as a relatively high density bubble that is crossing the lower density regions of the outer corona. This is clearly seen in Figure~\ref{fig:Magnetosphere}. One can see that in the close-orbit case, the magnetosphere density is comparable with that of the ambient corona, while in the longer-orbit case, the magnetosphere represents a bubble of higher density than that of the ambient corona. As a result, in the short-orbit case, the modulation of the radio flux is mostly due to the magnetosphere-corona interaction since the magnetosphere replace the ambient corona with an overall similar density. In contrast, in the longer-orbit case, the modulations are due to the density structure of the magnetosphere itself, since the magnetosphere replaces the corona with larger density along the path of the radio waves.

Figure \ref{fig:evol-long} shows the density modulations in the ambient medium due to the planetary motion in the long orbit scenario. Similar to the short orbit case (Figure \ref{fig:evol-short}), the telescope observing point and the LOS are marked by the thick black (Y=0) line. Figure \ref{fig:evol-long} b indicates that planetary magnetosphere modulates the density near the line of sight (thick black line) much before the mid-transit point. Figure \ref{fig:evol-long} d-f indicate that density modulation due to the planetary magnetosphere along the line of sight remains even after the completion of transit. 

Figure \ref{fig:exo_long_transit1} shows the synthetic light curves of the radio flux at different frequencies as a function of the orbital phase when planet is orbiting further from the star. The figure shows the modulations for the two different planetary magnetic field strengths.

Figure \ref{fig:exo_long_transit1} shows significant drop in the radio intensity value at the mid transit phase. We notice very significant (almost 45~\%) drop in the radio intensity value for the 100 MHz frequency; drop is much lower (around 5-10~\%) for other frequencies (250 MHz-1 GHz). We also notice slight increase in the radio intensity just before and after the transit similar like short-orbit scenario. This is because magnetosphere refract the radio waves towards the observer just before and after the transit. Transit behaviour of the 30~MHz radio frequency also shows a drop in the radio intensity value at the mid transit and a increase in the value of radio intensity just before and after the transit. 

We find quite different behaviour of radio transit for very low frequency (see Figure \ref{fig:exo_long_transit1}). Between phase of 0.4 and mid-transit, the radio flux increases by 10~\%, and then drops slightly around mid transit. Between phase of 0.52 and 0.6, we again find an increase, and then drops around the phase 0.6. This is a clear, and a very extensive focusing effect of the stellar radio emission by the moving planetary ``bubble". We notice some modulation in the radio intensity value around phase of 0.15 and 0.85 for all radio frequencies. Looking at Figure~\ref{fig:exo_long_transit1}, these phases seemed to be the locations of the dense stellar helmet streamers. This modulation is probably the effect of the planet crossing the streamers in the outer corona.  

In all scenarios, we notice a very clear difference in the magnitude of the modulations between the strong and weak field cases, and a slight difference in phase. Variations between the compressed day side magnetopause and the stretched magnetotail seem not to be visible in the radio lightcurves when the planet is a orbiting the star at a greater distance and interacting with the outer corona.

\section{Discussion}
\label{Discussion}

\subsection{Exoplanet Magnetic Field from the Radio Transit}
\label{exo-mag}

Previous studies assumed that the exoplanetary magnetosphere is a source for an intense non-thermal radio emission. Interaction between the stellar wind and the planetary magnetosphere causes long term variation in the magnetospheric radio emission \citep{desc83}. The magnetospheric radio emission is directly proportional to the stellar wind energy input at the planetary magnetosphere. It has been shown that solar wind energy input at the planetary magnetosphere depends on the solar wind-magnetosphere stand off distance \citep{desc84}. As solar wind-magnetosphere stand off distance depends on the planetary magnetic field strength, thus one can easily estimate the planetary magnetic field from the scaling relationship between the solar wind-magnetosphere stand off distance and the magnetospheric radio emission \citep{desc84, mill88, farr99, zark01a, lazi04}. However, auroral radio emission from exoplanets seem to be below the observable threshold \citep[e.g.,][]{Burkhart2017,lync18}. Here, we are proposing an alternative approach for calculating exoplanet magnetic field from the radio transit.

Instead of detecting planetary magnetosphere as a source of radio emission, here we focus on the modulation in the radio intensity from the host star caused by the planetary magnetosphere to characterize the exoplanetary magnetic field. Specifically, we aim to figure out some scaling relationship between the radio intensity modulation during the transit and the planetary magnetic field. For this purpose, we consider six different scenarios with three different planetary magnetic field strengths, namely 0.33 G, 1 G and 3 G respectively and two different orbital distances, namely 10 R$_\star$ and 20 R$_\star$ (the intermediate case of 1 G planetary field strength is not shown in details in the paper). 

\begin{figure*}[!ht]
\centering
\begin{tabular}{cc}
\includegraphics*[width=\linewidth]{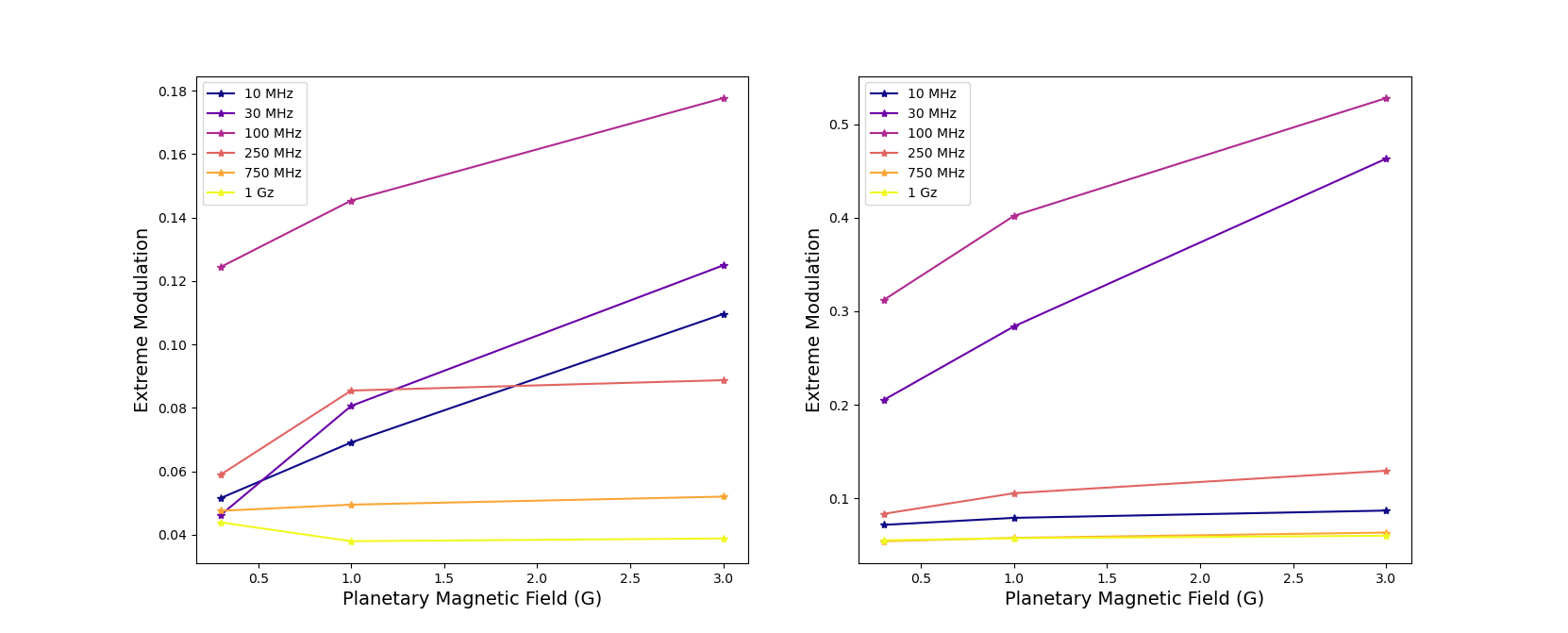}
\end{tabular} 
\caption{Left Panel: Relationship between Extreme Modulation and exoplanet magnetic field strength for short orbit scenario. Right Panel: Relationship between Extreme Modulation and exoplanet magnetic field strength for long orbit scenario. Extreme Modulation is defined as the difference between the maximum and minimum radio intensity values during the transit.}
\label{fig:exo-mag-field}
\end{figure*}

In order to get the simplest scaling of the radio flux modulation with the planetary field strength, We define the "Extreme Modulation" as the absolute difference between the maximum and minimum radio intensity value during the transit. Please note that we calculate the Extreme Modulation using the normalized radio transit dataset for each frequency. Figure~\ref{fig:exo-mag-field} shows the relationship between the Extreme Modulation and the planetary magnetic field strength. In the long orbit scenario, we find a clear increase in the Extreme Modulation value with the planetary magnetic field strength for all radio frequencies (Right panel of Figure~\ref{fig:exo-mag-field}). In the short orbit scenario, we also find a moderate increase in the Extreme Modulation value with the planetary magnetic field, but for some frequencies, it drops first and then increases again (Left panel of Figure~\ref{fig:exo-mag-field}). Please note that our data sets consist of only three data points, which is insufficient in order to derive a scaling law between the Extreme Modulation and the planetary magnetic field strength. For that, many simulations considering different planetary magnetic field strengths, star-planet distance and stellar magnetic map are needed. We also note that the size of the planetary magnetosphere is larger for higher planetary magnetic field (see Figure~\ref{fig:Magnetosphere}). Our initial results are very promising, indicating the possibility of exoplanet magnetic field determination from the radio transit.

\begin{figure*}
\begin{minipage}{.5\textwidth}
  \centering
  % include first image
  \includegraphics[width=\linewidth]{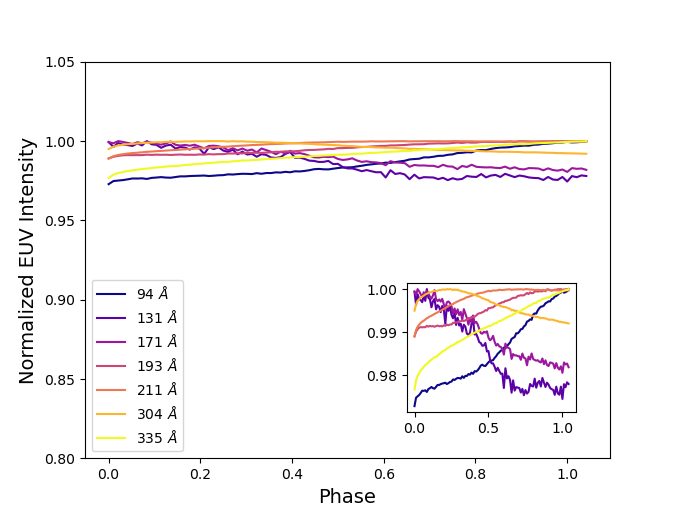}  
%  \caption{}
%  \label{fig:sub-firsteu}
\end{minipage}
\begin{minipage}{.45\textwidth}
  \centering
  % include second image
  \includegraphics[width=\linewidth]{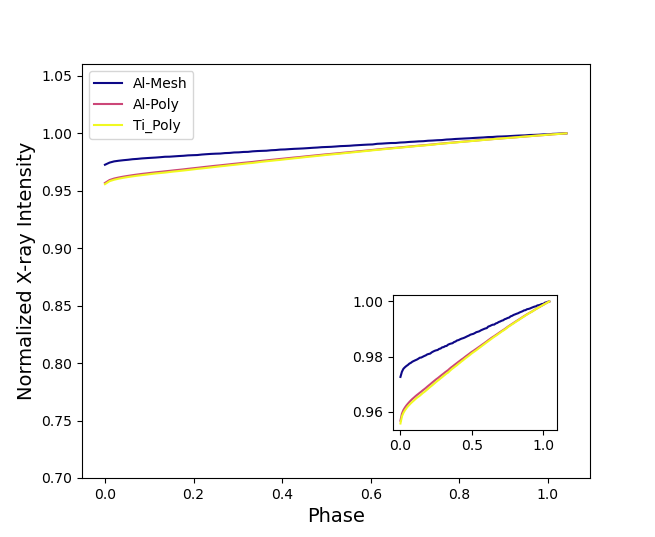}  
%  \caption{}
%  \label{fig:sub-secondxray}
\end{minipage}
\caption{Left Panel: Synthetic light curves for different EUV frequencies. Right Panel: Synthetic light curves for different X-ray frequencies. Al-mesh, Al-poly and Ti-poly corresponds to different X-ray filter.}
\label{fig:euv-xray}
\end{figure*}

\subsection{Exoplanets in EUV and X-ray:}
\label{uv-xray}
Most of the magnetically active stars are known to produce emissions in the EUV and X-ray bands, which are well-observed. It is possible that the planetary motions can modulate the EUV and X-ray emission from the host star as well. Since our model can also produce synthetic X-ray and EUV images \citep[see full description in, e.g.,][]{vand14,sach19}, we compare the synthetic modulations produced in these band to those obtained in the radio bands.

Left panel of the Figure~\ref{fig:euv-xray} shows the synthetic light curves for different EUV frequencies as a function of the orbital phase. Right panel of the Figure~\ref{fig:euv-xray} also shows the same but for the X-ray frequencies. In both cases, we do not find any significant modulation. There is some 3-4 \% modulation (see Zoomed in portion) but that is not because of exoplanet. That is probably due to the fact that our star-planet interaction model is time dependent; solutions of our model are not always perfectly steady, and we believe that these modulations are due to the non-steadiness of the stellar solution, and not due to the planetary motion.

One can expect some dip in the mid-transit phase as the planet is supposed to block the incoming X-ray and EUV emission. However, we do not find any such characteristic dip in the X-ray and EUV transit spectrum. First, the low-resolution magnetogram we use for HD 189733 produces very small amount of X-ray and EUV emission since no large active regions are included. Additionally, the orbital period of the planet is only two days in our simulation and the planet remains in the transit phase for very short time (only four to five hours). Please note that we have not considered the presence of large starspots and flares in our simulation setup. Large starspots and flares are know to produce high amount of EUV and X-rays; thus presence of large starspots and flares may increase the total X-ray and EUV flux value in the observable range. It may be then possible to observe exoplanet transits in EUV and X-ray \citep[See][]{popp13}.

\subsection{The Potential of Exoplanets Radio Transits Observations and Simulations}
\label{FutureofRadioTransits}

When an exoplanet transits its host star, stellar emission is expected to be absorbed or scattered by the planetary atmosphere or magnetosphere. One can use the resulting transit spectrum to characterize the atmosphere of that exoplanet. This techniques has been used successfully to characterize the atmosphere of many hot Jupiter, mini Neptune and super Earth like exoplanets \citep{seag20, vida03, krei14, knut14}. Host star can emit in different bands like visible, radio, EUV etc. During the propagation through the stellar or planetary atmosphere, these emitted waves are refracted (bent) in response to the atmospheric index of refraction gradient. This process is important as it modifies the atmospheric path traversed by the emitted waves, eventually impacting the collection of emitted waves by the observing telescope thus the transit. Previous studies have explored the effect of refraction on an exoplanet light curve/ transit spectrum \citep{hui02, sidi10, garc12, misr14}. Significant impact of refraction has been observed in our solar system during lunar eclipse and the 2004 Venus transit \citep{pasa11, garc12}.

In principle, it is very difficult to determine from where the refracted rays are reaching the observing telescope. However, we can think of some simple tentative scenarios. We first think the scenario when the planet is far away from the mid-transit LOS. In this case, although the moving planet modulates the density surrounding its position, it is very less likely that refracted rays will be able to reach the telescope. Rays that are coming from the star are only refracted by the non-modulated density variations of the stellar corona. One can expect almost no modulation or very less modulation in the transit spectrum. However, density modulation due to planetary motion does affect the refraction pattern significantly. When the planet gets closer to the LOS, just before the beginning of the transit, one can expect a significant impact of refraction on the transit spectrum. In this scenario, more refracted rays are likely to reach the telescope due to the strong density variations between the coronal and magnetospheric plasma, especially at the compressed region near the magnetopause. Thus one can expect an increase in flux just before the transit. Next, we consider the situation when the planet is at mid-transit. In this scenario, the planet and the planetary magnetosphere are supposed to block a significant amount of the incoming rays, while some refracted rays still reach the telescope. One can expect a dip in the observed flux. Finally, we consider the situation when the planet just completes the transit and moves out of transit. More refracted rays in addition to direct incoming rays are also reaching the telescope, due to the crossing of the LOS by the other side of the magnetosphere/magnetopause. Thus one can also expect an increase in the flux just after the transit. A general increase in the observed flux just before and after the transit, and a dip during the transit is mainly due to atmospheric lensing and it is described in \cite{sidi10}. Our results for radio waves refraction are overall consistent with these scenarios, while they are more detailed due to the non-ideal setting we choose to use here.

The structure of the stellar corona generally consists of closed magnetic loop (helmet streamers) and regions where the magnetic field lines are open to space. The so-called Alfv\'en surface defines the surface where the solar wind speed is equal to the local Alfv\'en speed. We generally find a slow, more dense wind structure near the top of the helmet streamer, and a faster, less dense wind away from these regions. Thus, the topology of the stellar magnetic field influences the coronal structure significantly (see \cite[e.g.,][]{McComas2007,reve15, stru15, perr21, hazr21}). The dense, hot closed field regions are actually the major source region for the background stellar coronal radio emission. Specifically, strong magnetic field near active regions is the source for the highest frequency radio emission \citep{mosc18}. Therefore, the controlling factor for the stellar radio emission (excluding transient radio bursts) is the stellar surface magnetic field structure \citep{vare18}. In our model, the low-resolution ZDI map for HD 189733 represents a relatively simpler magnetic structure (almost dipolar) with helmet streamer regions located mainly near the equator. Since the low-resolution magnetogram we use does not include active regions, it is likely that the actual radio flux from HD 189733, especially at higher frequencies is larger than the one modeled here. 

Our simulations show clear trends in the modulations of the radio flux for different frequencies, and planetary orbital distance from the star. Moreover, they show notable differences in the modulations as a function of the magnetic field strength. Interestingly, even the non-magnetized case has shown some modulations due to its induced magnetosphere. Our simulations demonstrate that both magnitude and phase differences exist in the radio flux modulations patterns, and that these differences could potentially be related to the planetary field strength. 

Despite of the fact that the radio flux from most planet-hosting stars is weak \cite[there are some potential selected targets, see][]{cohe18}, and that radio signals are very noisy, our simulations show that the planetary modulations of the radio signal could cover significant part of the phase curve due to the extended impact of the planetary magnetosphere. A combination of radio observations in both low and high radio frequencies, and a detailed modeling of the background stellar corona (driven by magnetic observations of the star) could provide a promising way to characterize and constrain the planetary magnetic field. Such a characterization could significantly improve our understanding about the internal structure of exoplanets, as well as their atmospheric evolution \citep{Gronoff2020}. 

Here, we choose to simulate a real system in order to capture a-symmetries in our solutions. We show that even in a non-uniform, more realistic case, radio modulations are clear and visible. However, idealized cases may help to better characterize the shape, magnitude, and phase location of different modulations as a function of magnetic field strength, orbital phase, and perhaps spectral type/stellar field strength. Specifically, if shape, magnitude, and phase location could be generalized, Machine Learning (ML) techniques could be adopted to search for planetary radio modulation in the large, available data sets. Such an approach has been adopted to detect exoplanets in other types of exoplanets data sets \cite[e.g.,][]{Malik2022}.

Previous studies already indicated the possibility of detecting the signature of exoplanet transit around the host star using upcoming SKA in the low frequency range \citep{pope19} and ALMA in the high frequency range \citep{selho13, selho20}. However, radio flux from the planet hosting star should be sufficiently strong to be observable; one may use few known stars with observable radio flux for that purpose \citep[See][]{wend95, slee03, etangs11, villa14, fich17, moha21}. Many of these previous studies placed an upper limit on the flux density and mass loss rates on the wind of host stars. \cite{villa14} detected all four selected stars in their study in the Ka band (centre frequency 34.5 GHz) using Very Large Array (VLA) and only able to put an upper limit on the flux density for other frequencies. The upgraded version of the existing VLA \citep[ngVLA;][]{oste18} will have very high sensitivity and can detect few observable stars in the radio band. However, we note that most of the stars are detected in non-thermal radio emission, especially in flares. In our study, we only consider the thermal radio emission, which is difficult to detect with present radio instruments. We may need to wait for more sensitive radio interferometers to detect thermal radio emission from planet-hosting stars.

Our next studies will be dedicated to perform a grid of idealized models of the star and the planet using dipole fields for both. In addition to the planetary field strength and orbital separation, we plan to investigate the radio modulations as a function of planetary field polarity (with respect to the stellar field), and planetary inclination. Moreover, we plan to investigate the effect of small-scale active regions on the radio modulations, and on the overall radio flux. We will either impose artificial active regions or alternatively, we will use solar magnetograms that include these small-scale features. 

\section{Conclusion}
\label{Conclusion}

Observing and characterizing magnetic fields of exoplanets are important for the understanding of their internal structure and atmospheric evolution. In this study, we perform a time-dependent star-planet interaction (SPI) simulation to study the same. We use HD 189733 as a central star and one hot Jupiter planet is moving around that star. We use a set of simulations aiming to demonstrate the feasibility of observing exoplanetary magnetic field using radio transit observations. 

Our simulations show some clear repeated trends in all radio frequencies, as well as some differences in these trends between the low- and high-range of frequencies. Moreover, our simulations demonstrate a clear dependence of the modulations on magnetic field strength, in terms of magnitude of the modulations, and a phase-dependence of some modulation features. Thus, our simulations suggest that the magnitude of the exoplanetary field could potentially determined from radio transit observations.

Our initial study here provides a solid background for an extended parametrization of the transit modulations of radio emissions from the star. Future work should combine simulations of the stellar corona and the exoplanet, and radio observations of stars with a feasibly observable radio flux. The former would provide specific features in the data that indicate the planetary strength, while the latter would provide the datasets in which those features may appear. ML tools would be ideal for this task. Future more sensitive radio interferometers may help us to detect these kinds of radio modulations in the planetary transit.

Additionally, we note that our study does not consider any kind of stellar magnetic variability. Our model only focuses on the thermal radio emission, not the coherent (non-thermal) radio emission. Observations indicate significant increase in the stellar X-ray and radio emission during flare and coronal mass ejection (CME). Effective cleaning of stellar magnetic variability (flare and CME effect) from the Radio transit spectrum is necessary to understand the impact of exoplanet.

\begin{acknowledgments}

This work is supported by NASA grant 80NSSC20K0840. Simulation results were obtained using the (open source) Space Weather Modeling Framework, developed by the Center for Space Environment Modeling, at the University of Michigan with funding support from NASA ESS, NASA ESTO-CT, NSF KDI, and DoD MURI. We also thank Dibyendu Nandi for the discussion and suggestions. The simulations were performed on NASA's Pleiades cluster under SMD-20-52848317.
\end{acknowledgments}

%\bibliographystyle{yahapj}
%\bibliography{reference_exo}

\end{document}